\documentclass[twocolumn,showpacs,amsmath,amssymb,prl,nofootinbib]{revtex4}
\usepackage{graphicx}
\usepackage{dcolumn}
\usepackage{bm}
\usepackage{psfig}

\newcommand{\ppbar}{p \overline{p}}
\newcommand{\pbar}{\overline{p}}
\newcommand{\Lambar}{\overline{\Lambda}}
\newcommand{\jpsi}{J/\psi}
\newcommand{\pip}{\pi^+}
\newcommand{\pin}{\pi^-}
\newcommand{\pio}{\pi^0}

\newcommand{\etap}{\eta^{\prime}}
\newcommand{\g}{\gamma}
\newcommand{\ar}{\rightarrow}
\newcommand{\kp}{K^+}
\newcommand{\kn}{K^-}

\begin{document}

\title{ \bf \boldmath Observation of a resonance $X(1835)$ in 
$\jpsi\ar\g\pip\pin\etap$ }

\author{M.~Ablikim$^{1}$, J.~Z.~Bai$^{1}$, Y.~Ban$^{11}$, J.~G.~Bian$^{1}$, 
X.~Cai$^{1}$, H.~F.~Chen$^{16}$, H.~S.~Chen$^{1}$, H.~X.~Chen$^{1}$, 
J.~C.~Chen$^{1}$, Jin~Chen$^{1}$
% ...
\vspace{0.2cm} \\(BES Collaboration)\\
\vspace{0.2cm}
{\it $^{1}$ Institute of High Energy Physics, Beijing 100049, People's Republic of China\\
% ...
$^{11}$ Peking University, Beijing 100871, People's Republic of China\\
% ...
$^{16}$ University of Science and Technology of China, Hefei 230026, People's Republic of China\\
% ...
}}

\date{\today}

\begin{abstract}
{ The decay channel $\jpsi\ar\g\pip\pin\etap$ is analyzed using a sample of 
$5.8\times 10^7$ $\jpsi$ events collected with the BESII detector. A resonance,
the $X(1835)$, is observed in the $\pip\pin\etap$ invariant mass spectrum with
a statistical significance of 7.7$\sigma$.  A fit with a Breit-Wigner function
yields a mass $M=1833.7\pm6.1(stat)\pm2.7(syst)$ MeV/c$^2$, a width  
$\Gamma=67.7\pm20.3(stat)\pm7.7(syst)$ MeV/c$^2$ and a product branching 
fraction $B(\jpsi\ar \g X)\cdot B(X\ar \pip\pin\etap)$ = $(2.2 \pm 0.4(stat)
\pm0.4(syst)) \times 10^{-4}$. The mass and width of the $X(1835)$ are not 
compatible with any known meson resonance.  Its properties are consistent with
expectations for the state that produces the strong $p\bar{p}$ mass threshold 
enhancement observed in the $\jpsi\ar\g p \bar{p}$ process at BESII. }
\end{abstract}
\maketitle


An anomalous enhancement near the mass threshold in the $p\pbar$ invariant mass
spectrum from $\jpsi\rightarrow\gamma p\pbar$ decays was reported by the BES II
experiment~\cite{gpp}. This enhancement was fitted with a sub-threshold 
$S$-wave Breit-Wigner resonance function with a mass $M=1859^{+~3+~5}_{-10-25}$
~MeV/c$^2$, a width $\Gamma<30$~MeV/c$^2$ (at the $90\%$ C.L.) and a product 
branching fraction (BF)  $B(\jpsi\ar\g X)\cdot B(X\ar\ppbar)$ = $(7.0\pm 0.4
(stat)^{+1.9}_{-0.8}(syst)) \times 10^{-5}$.
This surprising experimental observation has stimulated a number of theoretical
speculations~\cite{ppbar, theory, gao, yan, fsi1, fsi2} and motivated the 
subsequent experimental observation of a strong $p\Lambar$ mass threshold 
enhancement in $J/\psi \rightarrow p K^- \Lambar$ decay~\cite{pkl}. Among
various theoretical interpretations of the $p\pbar$ mass threshold enhancement,
the most intriguing one is that of a $p\pbar$ bound state, sometimes called 
{\it baryonium} ~\cite{ppbar, baryonium, yan}, which has
been the subject of many experimental searches~\cite{richard}.

The baryonium interpretation of the $p\pbar$ mass enhancement requires
a new resonance with a mass around 1.85~GeV/c$^2$ and it would be 
supported by the observation of the resonance in other decay channels.
Possible decay modes for a $p\pbar$ bound state, suggested in 
Ref.~\cite{gao,yan}, include $\pip\pin\etap$.  
In this letter, we report an analysis on the 
$\jpsi\ar\g\pip\pin\etap$ decay channel, and the observation of a resonance in 
the $\pip\pin\etap$ mass spectrum with a mass around 1835 ~MeV/c$^2$, where
the $\etap$ meson is detected in two decay modes, 
$\etap\ar\pip\pin \eta (\eta\ar\g\g)$ and $\etap\ar\g\rho$.
In the following, this resonance is designated as the $X(1835)$. 
The results reported here are based on a sample of $5.8 \times 10^7$ $J/\psi$ 
decays detected with the upgraded Beijing Spectrometer (BESII) at the Beijing 
Electron-Positron Collider (BEPC).


BESII is a large solid-angle magnetic spectrometer that is described in detail
in Ref.~\cite{BESII}. Charged particle momenta are determined with a resolution
of $\sigma_p/p = 1.78\%\sqrt{1+p^2(\mbox{\rm GeV/c}^2)}$ in a 40-layer 
cylindrical main drift chamber (MDC). Particle identification is accomplished 
by specific ionization ($dE/dx$) measurements in the MDC and time-of-flight 
(TOF) measurements in a barrel-like array of 48 scintillation counters. The 
$dE/dx$ resolution is $\sigma_{dE/dx} = 8.0\%$; the TOF resolution is measured
to be $\sigma_{TOF} = 180$~ps for Bhabha events. Outside of the time-of-flight
counters is a 12-radiation-length barrel shower counter (BSC) comprised of gas
tubes interleaved with lead sheets. The BSC measures the energies and 
directions of photons with resolutions of
$\sigma_E/E\simeq 21\%/\sqrt{E(\mbox{GeV})}$, $\sigma_{\phi} = 7.9$ mrad, 
and $\sigma_{z}$ = 2.3 cm. The iron flux return of the magnet is instrumented 
with three double layers of counters that are used to identify muons. In this 
analysis, a GEANT3-based Monte Carlo (MC) package with detailed consideration 
of the detector performance is used. The consistency between data and MC has 
been carefully checked in many high-purity physics channels, and the agreement
is reasonable~\cite{simbes}.

\begin{figure}[hbtp]
          \centerline{\psfig{file=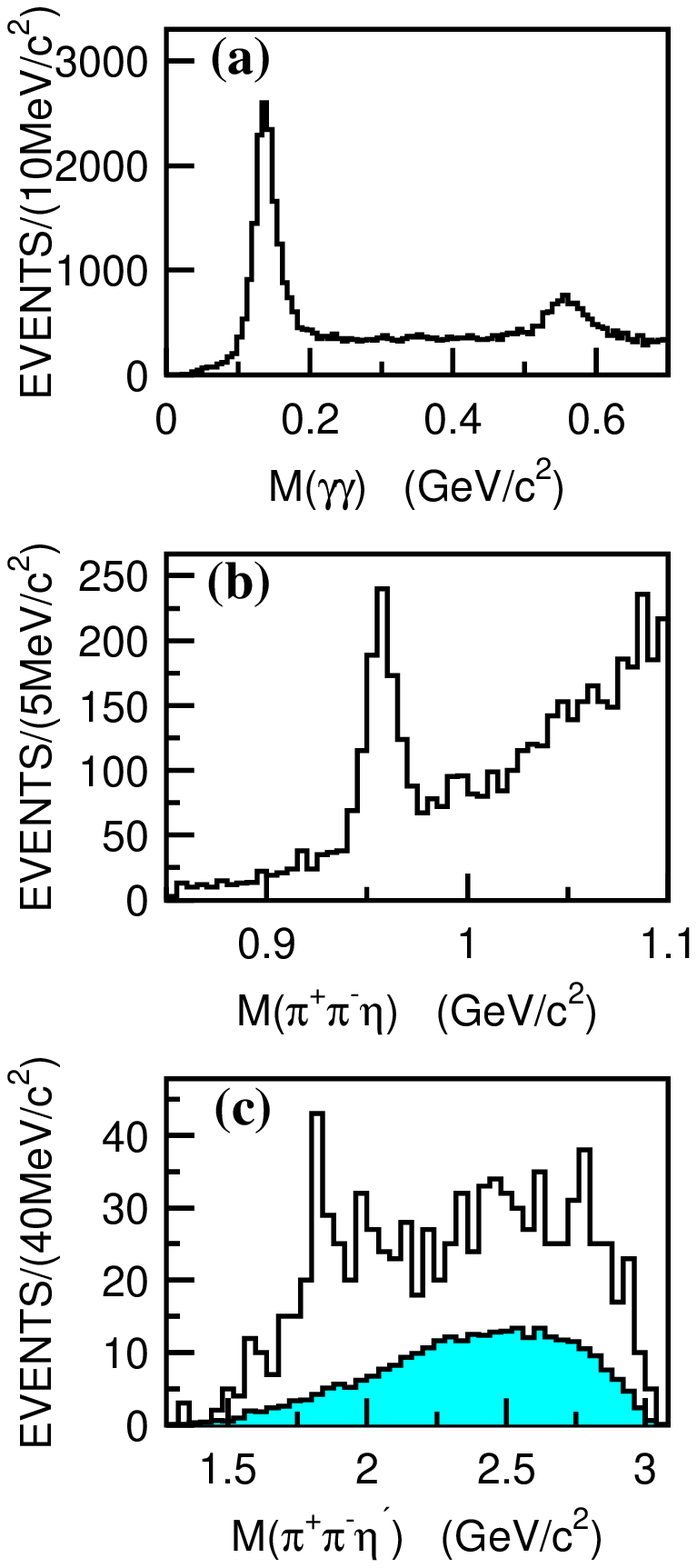,height=10.0cm,width=5.0cm}}
          \caption{ Invariant mass distributions for selected
           $\jpsi\ar\g\pip\pin\etap(\etap\ar\pip\pin\eta,\eta\ar\g\g)$ 
           candidate events.
           (a) The invariant mass distribution of $\g\g$ pairs.
	   (b) The $\pip\pin\eta$ invariant mass distribution. 
           (c) The $\pip\pin\etap$ invariant mass distributions; 
               the open histogram is data and the shaded histogram
               is $\jpsi\ar\g\pip\pin\etap$
               phase-space MC events (with arbitrary normalization). }
	\label{etapipi}
	\end{figure}

For the $\jpsi\ar\g\pip\pin\etap(\etap\ar\pip\pin\eta,\eta\ar\g\g)$ channel,
candidate events are required to have four charged tracks, each of which is 
well fitted to a helix that is within the polar angle region 
$|\cos \theta|<0.8$ and with a transverse momentum larger than $70$~MeV/c. 
The total charge of the four tracks is required to be zero.  For each 
track, the TOF and $dE/dx$ information are combined to form particle 
identification confidence levels for the $\pi, K$ and $p$ hypotheses; the 
particle type with the highest confidence level is assigned to each track.  
At least three of the charged tracks are required to be identified as pions. 
Candidate photons are required to have an energy deposit in the BSC 
greater than 60~MeV and to be isolated from charged tracks by more than 
$5^{\circ}$; the number of photons are required to be three. A four-constraint
(4C) energy-momentum conservation kinematic fit is performed to the 
$\pip\pin\pip\pin\g\g\g$ hypothesis and the $\chi^{2}_{4C}$ is required to be 
less than 8 and also less than the $\chi^2$ for the kinematically similar 
$\kp\kn\pip\pin\g\g\g$ hypothesis.  An $\eta$ signal is evident in the 
$\g\g$ invariant-mass distribution of all $\g\g$ pairings 
(Fig.~\ref{etapipi}(a)). In order to reduce combinatorial backgrounds from 
$\pio\ar\g\g$ decays, we require that the invariant masses of all $\g\g$ 
pairings are greater than 0.22 GeV/c$^2$. Candidate $\eta$ mesons are selected
by requiring $|M_{\g\g}-m_{\eta}|<0.05$ GeV/c$^2$. The events are then 
subjected to a five-constraint (5C) fit where the invariant mass of the 
$\g\g$ pair associated with the $\eta$ is constrained to $m_{\eta}$, and 
$\chi^{2}_{5C}<15$ is required. The 5C-fit improves the 
$M_{\pip\pin\eta}$ mass resolution from 20 MeV/c$^2$ (for the 4C-fit) to 
7 MeV/c$^2$. Figure~\ref{etapipi}(b) shows the $\pip\pin\eta$ invariant 
mass distribution after the 5C fit, where a
clear $\etap$ signal is visible. For $\etap$ candidates, we select 
$\pip\pin\eta$ combinations with $|M_{\pip\pin\eta}-m_{\etap}|<0.015$ 
GeV/c$^2$. In a small fraction of events, more than one combination passes the 
above selection. In these cases the combination with $M_{\pip\pin\eta}$ 
closest to $\etap$ mass is used~\cite{optimum}. The  $\pip\pin\etap$ invariant
mass spectrum for the selected events is shown in Fig.~\ref{etapipi}(c), 
where a peak at a mass around 1835 MeV/c$^2$ is observed.

\begin{figure}[hbtp]
          \centerline{\psfig{file=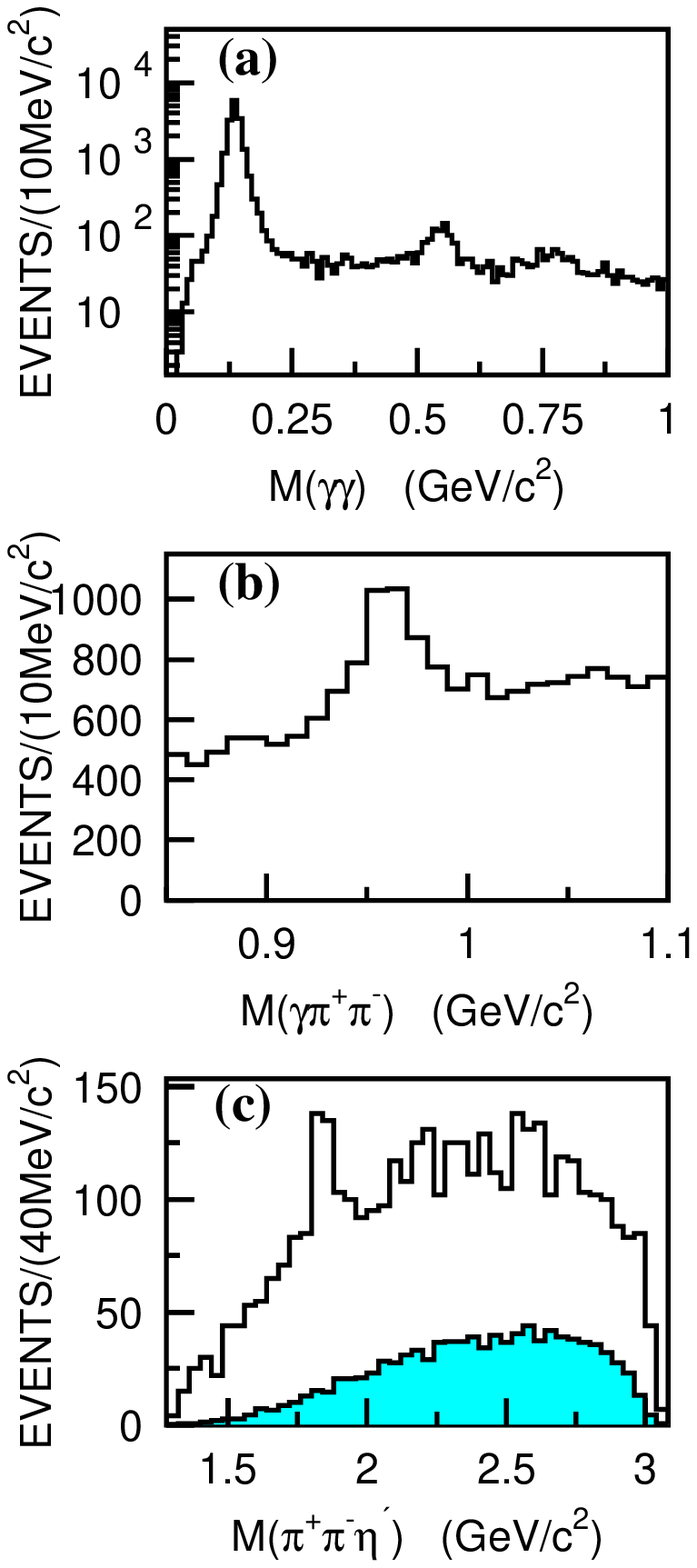,height=10.0cm,width=5.0cm}}
          \caption{ Invariant mass distributions for the selected
           $\jpsi\ar\pip\pin\etap(\etap\ar\g\rho)$ candidate events,
           (a) The invariant mass distribution for $\g\g$ pairs.
	   (b) The $\g\pip\pin$ invariant mass distribution. 
           (c) The $\pip\pin\etap$ invariant mass distributions: 
               the open histogram is data and the shaded histogram
               is from $\jpsi\ar\g\pip\pin\etap$
               phase-space MC events (with arbitrary normalization). }
	\label{grho}
	\end{figure}

For the $\jpsi\ar\g\pip\pin\etap(\etap\ar\g\rho)$ channel, events with four 
charged tracks (with zero net charge) and two photons are selected. At least 
three of the charged tracks are required to be identified as pions. These 
events are subjected to a 4C kinematic fit to the $\pip\pin\pip\pin\g\g$ 
hypothesis, and the $\chi^{2}_{4C}$ is required to be less than 8 and less than
the $\chi^2$ for the $\kp\kn\pip\pin\g\g$ hypothesis. At this stage of the 
analysis, the primary remaining background contributions are due to
$\jpsi\ar\pip\pin\pip\pin\pio$, $\jpsi\ar\pip\pin\pip\pin\eta$, and
$\jpsi\ar\omega(\omega\ar\g\pio)\pip\pin\pip\pin$; these produce peaks at 
$m_{\pio}$, $m_{\eta}$ and $m_{\omega}$ in the $\g\g$ invariant mass 
distribution shown in Fig.~\ref{grho}(a). We suppress these backgrounds by 
rejecting events with $M_{\g\g}<0.22$ GeV/c$^2$, $|M_{\g\g}-m_{\eta}|<$0.05 
GeV/c$^2$ or 0.72 GeV/c$^2$$<M_{\g\g}<$ 0.82 GeV/c$^2$. To select  $\rho$ and 
$\etap$ signals, all $\pip\pin$ and $\g\pip\pin$ combinations are considered. 
The $\g\pip\pin$ invariant mass distribution shows an $\etap$ signal 
(Fig.~\ref{grho}(b)). We require that $|M_{\pip\pin}-m_{\rho}|<$~0.2~GeV/c$^2$
and $|M_{\g\pip\pin}-m_{\etap}|<0.025$~GeV/c$^2$. If more than one combination
passes these criteria, the combination with $M_{\g\pip\pin}$ closest to 
$m_{\etap}$ is selected~\cite{optimum}. For this channel there is also a
distinct peak near 1835~MeV/c$^2$ in the $\pip\pin\etap$ invariant mass 
spectrum (Fig.~\ref{grho}(c)).

To ensure that the peak near 1835~MeV/c$^2$ is not due to background, we have 
made extensive studies of potential background processes using both data and 
MC.  Non-$\etap$ processes are studied with $\etap$ mass-sideband events. The 
main background channel, $\jpsi\ar\pio\pip\pin\etap$, and other background 
processes with multi-photons and/or with kaons are reconstructed with the data.
In addition, we also checked for possible backgrounds with a MC sample of 60 
million $J/\psi$ decays generated by the LUND model~\cite{chenjc}. None of 
these background sources produce a peak around 1835 MeV/c$^2$ in the 
$\pip\pin\etap$ invariant mass spectrum.

\begin{figure}[hbtp]
          \centerline{\psfig{file=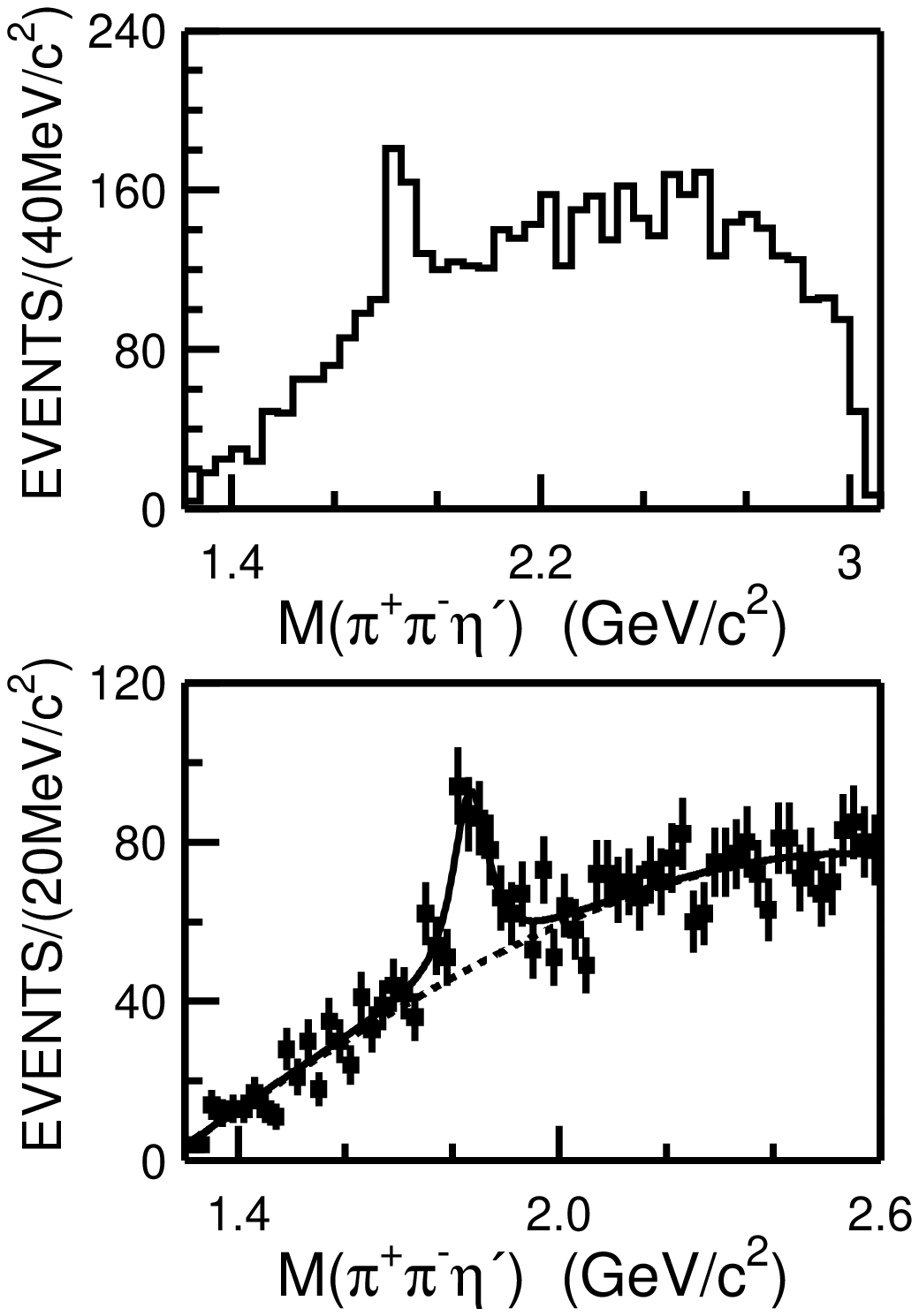,height=7cm,width=5.3cm}}
          \caption{ The $\pip\pin\etap$ invariant mass distribution for
           selected events from both the
           $\jpsi\ar\g\pip\pin\etap(\etap\ar\pip\pin\eta,\eta\ar\g\g)$
           and $\jpsi\ar\g\pip\pin\etap(\etap\ar\g\rho)$ analyses. The bottom
           panel shows the fit (solid curve) to the data (points with error 
           bars); the dashed curve indicates the background function. }
	\label{sum}
	\end{figure}

Figure~\ref{sum} shows the $\pip\pin\etap$ invariant mass spectrum for the 
combined $\jpsi\ar\g\pip\pin\etap(\etap\ar\pip\pin\eta)$ and 
$\jpsi\ar\g\pip\pin\etap(\etap\ar\g\rho)$ samples ($i.e.,$ the sum of the 
histograms in Figs.~\ref{etapipi}(c) and~\ref{grho}(c)). This spectrum is 
fitted with a Breit-Wigner (BW) function convolved with a Gaussian mass 
resolution function (with $\sigma = 13$~MeV/c$^2$) to represent the $X(1835)$ 
signal plus a smooth polynomial background function. The mass and width 
obtained from the fit (shown in the bottom panel of Fig.~\ref{sum}) are  
$M=1833.7\pm 6.1$~MeV/c$^2$ and $\Gamma=67.7\pm 20.3$~MeV/c$^2$. The signal
yield from the fit is $264\pm 54$ events with a confidence level 
45.5$\%$ ( $\chi^2/d.o.f.$ = 57.6/57) and $-2\ln L=58.4$.  A fit to the mass 
spectrum without a BW signal function returns $-2\ln L = 126.5$. The change in
$-2\ln L$ with $\Delta(d.o.f.) = 3$  corresponds to a statistical significance
of 7.7~$\sigma$ for the signal.

Using MC-determined selection efficiencies of $3.72\%$ and $4.85\%$ for the
$\etap\ar\pip\pin\eta$ and $\etap\ar\g\rho$ modes, respectively, we determine 
a product BF of 
\begin{center}
$B(\jpsi\ar\g X(1835))\cdot B(X(1835)\ar\pip\pin\etap) = (2.2\pm 0.4)\times
10^{-4}.$

\end{center}

The consistency between the two $\etap$ decay modes is checked by fitting the 
distributions in Fig.~\ref{etapipi}(c) and Fig.~\ref{grho}(c) separately with
the method described above. The fit to Fig.~\ref{etapipi}(c) gives $M = 1827.4
\pm 8.1$ ~MeV/c$^2$ and $\Gamma = 54.2 \pm 34.5$~MeV/c$^2$ with a statistical
significance of 5.1$\sigma$. From the $68 \pm 26$ signal events obtained from 
the fit, the product BF is $B(\jpsi\ar\g X(1835)) \cdot 
B(X(1835)\ar\pip\pin\etap) = (1.8\pm 0.7)\times 10^{-4}$. Similar results are 
obtained if we only apply 4C kinematic fit in this analysis. For the fit to 
Fig.~\ref{grho}(c), the mass and width are determined to be $M = 1836.3\pm 
7.9$~MeV/c$^2$ and $\Gamma = 70.3\pm23.1$~MeV/c$^2$ with a statistical 
significance of 6.0 $\sigma$. For this mode alone, the signal yield of 
$193\pm 43$ signal events correspond to $B(\jpsi\ar\g X(1835))\cdot 
B(X(1835)\ar\pip\pin\etap)= (2.3\pm 0.5)\times 10^{-4}$.   The $X(1835)$ mass,
width and product BF values determined from  the two $\etap$ 
decay modes separately are in good agreement with each other.


The systematic uncertainties on the mass and width are determined by varying 
the functional form used to represent the background,  the fitting range of the
mass spectrum, the mass calibration, and possible biases due to the fitting 
procedure.  The latter are estimated from  differences between the input and 
output mass and width values from MC studies. The total systematic errors on 
the mass and width are 2.7 MeV/c$^2$ and 7.7~MeV/c$^2$, respectively. The 
systematic error on the branching fraction measurement mainly comes from the 
uncertainties of MDC simulation (including systematic uncertainties of the 
tracking efficiency and the kinematic fits), the photon detection efficiency, 
the particle identification efficiency, the $\etap$ decay branching fractions 
to $\pip\pin\eta$ and $\g\rho$, the background function parameterization, 
the fitting range of the mass spectrum, the requirements on numbers of photons,
the invariant mass distributions of $\g\g$ pairs in the two analyses, the 
$\pip\pin$ invariant mass distribution in $\etap\ar\g\pip\pin$ decays, MC 
statistics, the total number of $\jpsi$ events \cite{ssfang}, and the unknown 
spin-parity of the $X(1835)$. For the latter we use the difference between 
phase-space and a $J^{PC}=0^{-+}$ hypothesis for the $X(1835)$.  The total 
relative systematic error on the product branching fraction is $20.2\%$.


In summary, the decay channel $\jpsi\ar\g\pip\pin\etap$ is 
analyzed using two $\etap$ decay modes, $\etap\ar\pip\pin\eta$ and 
$\etap\ar\g\rho$. A resonance, the $X(1835)$, is observed with a high 
statistical significance of 7.7$\sigma$ in the $\pip\pin\etap$ invariant mass 
spectrum. From a fit with a Breit-Wigner function, the mass is determined to 
be $M = 1833.7\pm6.1(stat)\pm2.7(syst)$~MeV/c$^2$, the width is $\Gamma = 
67.7\pm20.3(stat)\pm7.7(syst)$~MeV/c$^2$, and the product branching fraction 
is $B(\jpsi\ar \g X)\cdot B(X\ar \pip\pin\etap)$ = 
$(2.2\pm0.4(stat)\pm0.4(syst)) \times 10^{-4}$. The mass and width of the 
$X(1835)$ are not compatible with any known meson resonance~\cite{pdg}. 
In Ref.~\cite{pdg}, the candidate closest in mass to the $X(1835)$ is 
the (unconfirmed) $2^{-+}$ $\eta_2(1870)$ with $M=1842\pm 8$~MeV/c$^2$. 
This state's width, $\Gamma = 225\pm 14$~MeV/c$^2$, is considerably larger 
than that of the $X(1835)$ (see also~\cite{BESI}, where the $2^{-+}$
component in the $\eta\pi\pi$ mode of $\jpsi$ radiative decay has
a mass $1840 \pm 15$~MeV/c$^2$ and a width $170\pm 40$~MeV/c$^2$).

We examined the possibility that the $X(1835)$ is responsible for the 
$p\bar{p}$ mass threshold enhancement observed in radiative
$\jpsi\ar\g p \bar{p}$ decays \cite{gpp}.  It has been pointed out that the 
$S$-wave BW function used for the fit in Ref.~\cite{gpp} should be modified to 
include the effect of final-state-interactions (FSI) on the shape of the 
$p\bar{p}$ mass spectrum ~\cite{fsi1, fsi2}. Redoing the $S$-wave BW fit to 
the $p\bar{p}$ invariant mass spectrum of Ref.~\cite{gpp} including 
the zero Isospin, $S$-wave FSI factor of Ref.~\cite{fsi2},  yields a mass  
$M = 1831 \pm 7$~MeV/c$^2$ and a width $\Gamma < 153$~MeV/c$^2$ (at the 90$\%$ 
C.L.)~\cite{nosys}; these values are in good agreement with the mass and width 
of $X(1835)$ reported here.   Moreover, according to Ref.~\cite{yan}, the 
$\pi\pi\etap$ decay mode is expected to be strong for a $\ppbar$ bound state. 
Thus, the $X(1835)$ resonance is a prime candidate for the source of the 
$p\bar{p}$ mass threshold enhancement in $\jpsi\ar \g p\bar{p}$ process. 
In this case, the $J^{PC}$ and $I^G$ of the $X(1835)$ could only be $0^{-+}$ 
and $0^+$, which can be tested in future experiments.  Also in this context, 
the relative $\ppbar$ decay strength is quite  strong:
$B(X\ar \ppbar)/B(X\ar \pip\pin\etap)\sim 1/3$~\cite{fsi-bf}. 
Since decays to $\ppbar$ are kinematically allowed only for a small portion 
of the high-mass tail of the resonance and have very limited phase space, 
the large $\ppbar$ branching fraction implies an unusually strong coupling 
to $\ppbar$, as expected for a $\ppbar$ bound state~\cite{zhu}.  However, 
other possible interpretations of the $X(1835)$ that have no relation to the 
$\ppbar$ mass threshold enhancement are not excluded.

We thank many Chinese theorists for their helpful 
discussions and suggestions, and Dr. A. Sibirtsev {\sl  et al.} for
providing the FSI functions. The BES collaboration acknowledges the staff of 
BEPC for the excellent performance of the machine. This work is supported in 
part by the National Natural Science Foundation of China under contracts Nos. 
19991480, 10225524, 10225525, 10425523, the Chinese Academy of Sciences under 
contract No. KJ 95T-03, the 100 Talents Program of CAS under Contract Nos. 
U-11, U-24, U-25, and the Knowledge Innovation Project of CAS under Contract 
Nos. KJCX2-SW-N10, U-602, U-34 (IHEP); by the National Natural Science 
Foundation of China under Contract No. 10175060 (USTC); and by the Department 
of Energy under Contract No. DE-FG03-94ER40833 (U Hawaii).


\begin {thebibliography}{99}
\bibitem{optimum}
The selection criteria are optimized by maximizing 
$N_s/\sqrt{N_{tot}}$, where $N_s$ is the expected number of 
$\jpsi\ar\g\pip\pin\etap$ signal events and $N_{tot}$ is the total expected 
number of signal and background events.
\bibitem{nosys} Systematic uncertainties are not included in the
 error of the mass and the upper limit of the width.
\bibitem{fsi-bf} The product BF determined from the fit that 
  includes FSI effects on the $p\bar{p}$ mass spectrum is within
  the systematic errors of the result reported in Ref.~\cite{gpp}.
\end{thebibliography}
\end{document}